# Transition from Amplitude Death to Oscillation Death in Coupled Chua's Circuits

S. Chakraborty[*]
Physics Department, Bidhan Chandra College, Asansol- 713304, West Bengal, India

In this piece of work, an important transition from amplitude death (AD) to oscillation death (OD) state in coupled Chua's circuit has been explored for the first time. Here, mean field diffusively (MFD) coupled Chua's circuits are when additionally coupled through direct coupling, it exhibits the AD-OD transition phenomena. The effect of design parameters and coupling parameters are investigated in details. For both identical and non-identical systems the AD make transition to OD through symmetry breaking. These are investigated through numerical simulations and are established through hardware experiment. The present study includes a new dimension to versatility of Chua's circuit in the field of nonlinear dynamics.

PACS numbers: 05.45.Xt, 05.45.Ac, 05.45.Pq, 07.50.Ek

## I. INTRODUCTION

Recently, the study on the coupled system gets great importance. Coupled system shows variety of interesting complex collective phenomena, such as synchronization, dynamical hysteresis, phase flip, oscillation quenching etc. [1-7]. Oscillation quenching is a phenomenon where all the coupled system suppresses the oscillatory behavior of each other through their interaction. Such quenched states are of two types; Amplitude death (AD) and Oscillation death (OD) [8, 9]. In AD the coupled system converges to same steady states (homogeneous) and is mainly important to stabilize the system viz. in laser [10], neuronal systems [11], etc. But, in OD the system shows heterogeneity in their quenched states [12]. It has also find an enormous importance in the diverse field [7, 8, 12, 13].

Koseska et al. in [8] show that such AD and OD can simultaneously occur in coupled Stuart-Landau oscillators using diffusive coupling and reported the transition from AD to OD. After then several works have been reported showing such transition in different coupled system [9, 14-18]. In those works different coupling method like; mean field diffusive (MFD) coupling [9], time-delay [14], dynamic coupling [15], conjugate coupling [16, 17], diffusive and repulsive coupling [18] etc. have been used.

In this paper, we first show the AD-OD transition in Chua circuit using most simple circuit modification in MFD coupled Chua's circuit. Chua's circuit is one of the most simple and efficient circuits. It shows wide variety of nonlinear dynamics within its so simple structure and it can be easily implemented in experiments [19, 20]. So far literature is concern, a good amount of works have already been done on the coupled Chua's circuit using different coupling schemes [17, 21, 22]. But in all cases the coupled system converges to a common steady state showing AD in them. Sharma et al. already reported AD with mean field diffusion in Chua circuit in [22]. We explore the dynamics of such MFD coupled Chua's circuit when additionally coupled through a resistor. It is well known that Chua's circuit is a third order autonomous system and its dynamics is described by three state variables [20]. In our work, we coupled one of the state variables of Chua's circuits through MFD coupling and another state variable through a simple variable resistor (i.e. direct diffusively coupled). Interestingly, in such condition the system dynamic shows a transition from AD to OD depending upon the strength of direct coupling. The OD is appeared through symmetry breaking. Such dynamics are described through numerical simulation and also supported by a prototype hardware experiment. The paper has been organized as follows. The equations describing the system dynamics for such modified MFD coupled Chua' circuit has been formulated in section II. The occurrences of AD, OD and their transition in the system have been discussed in section III through numerical simulations. The impact of coupling and design parameters on the transition phenomenon have also been discussed here. In establishment of such dynamics we perform a experiment on a proto type hardware circuit using off-the-self ICs. The details of the experiments and the experimental results are described in section IV. Some concluding remarks are given in the Section V.

[*]Email: saumenbcc@gmail.com

## II. SYSTEM EQUATION FORMULATION

For the sake of completeness, at first we describe the salient features of conventional Chua's circuit and then proceed to formulate the state equations for modified MFD coupled Chua's circuit. Figure 1 gives the structure of conventional Chua's circuit consisting of two capacitors, an inductor, a resistor and a nonlinear resistor. Applying Kirchhoff's current laws to the various branch points of Fig.1 one can obtain the state equations that describes the dynamics of the conventional system as [20],

$$C_1 \frac{dv_1}{dt} = \frac{(v_2 - v_1)}{r} - f(v_1) \quad (1a)$$

$$C_2 \frac{dv_2}{dt} = \frac{(v_2 - v_1)}{r} + i_L \quad (1b)$$

$$L \frac{di_L}{dt} = -v_2 \quad (1c)$$

Here, $v_1$, $v_2$ and $i_L$ are the voltage across the capacitors $C_1$ and $C_2$ and the current through L, respectively as shown in Fig.1.

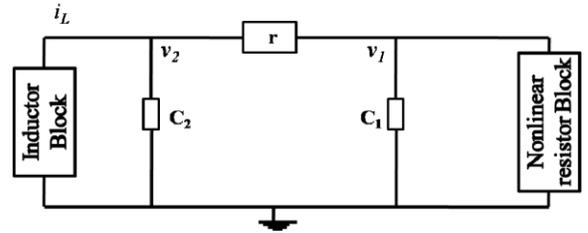

FIG.1. Block diagram of conventional Chua's Circuit.

$f(v_1)$ in equation (1a) represents the response of the nonlinear resistor and it can be mathematical represented as follows,

$$f(v_1) = m_0 v_1 + \left(\frac{1}{2}\right)(m_1 - m_0)[|v_1 + B_p| - |v_1 - B_p|] \quad (2)$$

Here, $m_0$ and $m_1$ are the inner and outer slopes of the typical characteristic curve of the Chua's diode. $B_p$ is the breakpoint of two slopes. The characteristic curve of the Chua's diode is shown in Fig. 2. Considering $\tau = \frac{t}{C_2 r}$, a dimensionless quantity, we derive three normalized state equations describing the dynamics of Chua's circuit in terms of three state variables as,

$$\frac{dx}{d\tau} = \alpha(y - x - f(x)) \tag{3a}$$

$$\frac{dy}{d\tau} = x - y + z \tag{3b}$$

$$\frac{dz}{d\tau} = -\beta y \tag{3c}$$

Where, $f(x) = m_0 r x + \left(\frac{1}{2}\right)(m_1 - m_0)r[|x+1| - |x-1|]$, $x = \frac{v_1}{B_p}$, $y = \frac{v_2}{B_p}$, $z = \frac{r i_L}{B_p}$, $\alpha = \frac{C_2}{C_1}$ and $\beta = \frac{C_2 r^2}{L}$.

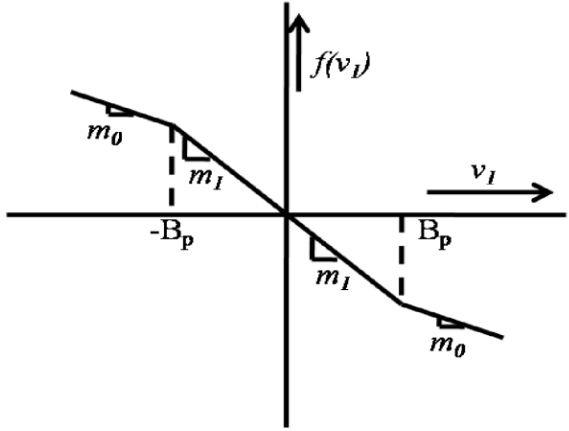

FIG.2. Variation of nonlinear function $f(v_1)$ with respect to $v_1$ as given in equation (2).

Now, we coupled two such circuits. The complete hardware circuit diagram of modified MFD coupled Chua's circuits is shown in Fig. 3. First, we connect the branch points having voltage $v_{12}$ and $v_{22}$ (i.e. $y$ state variable in normalized condition) through MFD coupling. The circuit equations of MFD coupled Chua's circuits are derived using Kirchhoff's current law and the equations are as follows,

$$C_{11}\frac{dv_{11}}{dt} = \frac{(v_{12}-v_{11})}{r_{11}} - f(v_{11}) \tag{4a}$$

$$C_{12}\frac{dv_{12}}{dt} = \frac{(v_{11}-v_{12})}{r_{11}} + i_{1L} + \left(\frac{1}{R_6}\right)\left(\frac{R_3}{R_1}\right)(v_{12} + v_{22}) - (v_{12}) \tag{4b}$$

$$L_1\frac{di_{1L}}{dt} = -v_{12} \tag{4c}$$

$$C_{21}\frac{dv_{21}}{dt} = \frac{(v_{22}-v_{21})}{r_{21}} - f(v_{21}) \tag{4d}$$

$$C_{22}\frac{dv_{22}}{dt} = \frac{(v_{21}-v_{22})}{r_{21}} + i_{2L} + \left(\frac{1}{R_6}\right)\left(\frac{R_3}{R_1}\right)(v_{12} + v_{22}) - (v_{22}) \tag{4e}$$

$$L_2\frac{di_{2L}}{dt} = -v_{22} \tag{4f}$$

Here, the first digit of subscript (i.e. 1 or 2) is used to represent first or second Chua's circuit, respectively. Now we connect the branch points having the voltage $v_{11}$ and $v_{21}$ (i.e. $x$ state variable in normalized condition) through a simple resistor $R_7$. Thus, the two Chua's circuits are direct diffusively coupled through these state variables. This is also shown in Fig.3. Then the above equation set are modified as given below,

$$C_{11}\frac{dv_{11}}{dt} = \frac{(v_{12}-v_{11})}{r_{11}} - f(v_{11}) + \frac{(v_{11}-v_{21})}{R_7} \tag{5a}$$

$$C_{12}\frac{dv_{12}}{dt} = \frac{(v_{11}-v_{12})}{r_{11}} + i_{1L} + \left(\frac{1}{R_6}\right)\left(\frac{R_3}{R_1}\right)(v_{12} + v_{22}) - (v_{12}) \tag{5b}$$

$$L_1\frac{di_{1L}}{dt} = -v_{12} \tag{5c}$$

$$C_{21}\frac{dv_{21}}{dt} = \frac{(v_{22}-v_{21})}{r_{21}} - f(v_{21}) - \frac{(v_{11}-v_{21})}{R_7} \tag{5d}$$

$$C_{22}\frac{dv_{22}}{dt} = \frac{(v_{21}-v_{22})}{r_{21}} + i_{2L} + \left(\frac{1}{R_6}\right)\left(\frac{R_3}{R_1}\right)(v_{12} + v_{22}) - (v_{22}) \tag{5e}$$

$$L_2\frac{di_{2L}}{dt} = -v_{22} \tag{5f}$$

Considering $\tau = \frac{t}{C_{12}r_{11}}$, we normalized the above equations and derive six normalized state equations describing the dynamics of such MFD as well as direct coupled Chua circuit as,

$$\frac{dx_1}{d\tau} = \alpha_1\{(y_1 - x_1 - f(x_1)) + g_1(x_1 - x_2)\} \quad (6a)$$

$$\frac{dy_1}{d\tau} = x_1 - y_1 + z_1 + d_1(m\left(\frac{y_1+y_2}{2}\right) - y_1) \quad (6b)$$

$$\frac{dz_1}{d\tau} = -\beta_1 y_1 \quad (6c)$$

$$\frac{dx_2}{d\tau} = \alpha_2\{(y_2 - x_2 - f(x_2)) - g_2(x_1 - x_2)\} \quad (6d)$$

$$\frac{dy_2}{d\tau} = \gamma[(x_2 - y_2 + z_2) + d_2(m\left(\frac{y_1+y_2}{2}\right) - y_2)] \quad (6e)$$

$$\frac{dz_2}{d\tau} = -\beta_2 y_2 \quad (6f)$$

Here, $\alpha_1 = \left(\frac{C_{12}}{C_{11}}\right)$, $\alpha_2 = \left(\frac{C_{12}r_{11}}{C_{21}r_{21}}\right)$, $\gamma = \left(\frac{C_{12}r_{11}}{C_{22}r_{21}}\right)$, $\beta_1 = \left(\frac{C_{12}r_{11}^2}{L_1}\right)$, $\beta_2 = \left(\frac{C_{12}r_{11}r_{21}}{L_2}\right)$, $f(x_1) = r_{11}f(v_{11}), f(x_2) = r_{21}f(v_{21})$. The factors $d$ [ $d_1 = \left(\frac{r_{11}}{R_6}\right)$ and $d_2 = \left(\frac{r_{21}}{R_6}\right)$ ], $m = \left(\frac{2R_3}{R_1}\right)$ and $g$ [ $g_1 = \frac{r_{11}}{R_7}$ and $g_2 = \frac{r_{21}}{R_7}$ ] signifies the diffusive strength, mean field strength and strength of direct coupling of the coupled system, respectively.

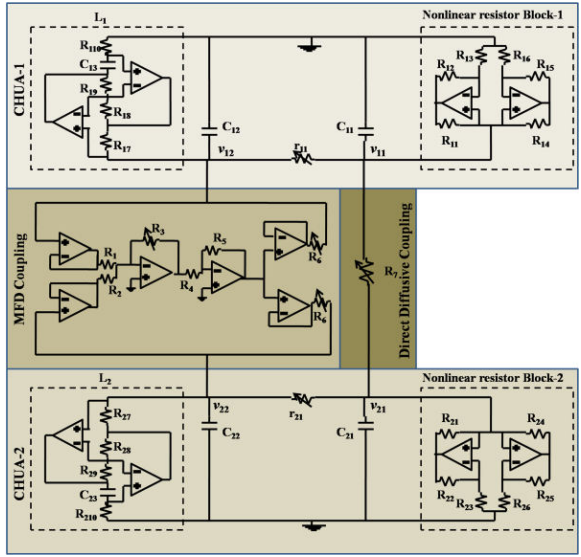

FIG.3. (Color online) Hardware circuit diagram for experiment of modified MFD coupled Chua's circuits.

## III. NUMERICAL SIMULATION RESULTS

The dynamics of MFD coupled as well as direct coupled Chua's circuit has been studied through numerical integration of the state equations given in (6) using the 4th order Runge-Kutta technique. Every time, to eliminate the initial transients' large amount of initial data is discarded. The influences of direct coupling on a MFD coupled Chua's circuit has been examined through simulation. The effect of $g$ parameter along with the conventional parameters like $\alpha, \beta, \gamma, d, m$ on the dynamics of the modified coupled Chua's circuit is obtained.

To make our observation simple we take the two systems identical, i.e. $\alpha_1 = \alpha_2 = \alpha$, $\beta_1 = \beta_2 = \beta$, $\gamma = 1$, $d_1 = d_2 = d$ and $g_1 = g_2 = g$. Initially, we set $g = 0$ that means the two circuits are under MFD coupling. In this condition for $\alpha = 10$, $\beta = 18.43$ the two isolated Chua's circuits shows chaotic oscillations. Now we set $d = 0.6$ and then gradually decrease $m$. It is observed that the chaotic dynamics of the coupled Chua's circuit would converge to the trivial homogeneous steady state i.e. AD occurs. The observation is depicted in Fig.4. Transition from oscillatory state to AD in parameter space $(d - m)$ for MFD coupled Chua's circuit is also simulated. The numerical simulation results show for higher value of mean field strength, higher value of diffusive strength is needed to make transition from oscillatory state to AD as investigated in [22] and this is shown in Fig.5. In Fig.5 beyond AD we observe the system oscillates with multi-period. But since we are interested only in quenching phenomena we simply defined them as oscillatory state (OS). When AD already occurs in the coupled system, we apply $g$ i.e. direct coupling between $x_1$ and $x_2$ takes place. With $g = 0.015$, the trivial steady state becomes unstable and give birth of two new inhomogeneous steady states through symmetry breaking (i.e. OD is created). Thus, we get a transition from AD to OD in modified coupled system. It is also observed that the difference between two inhomogeneous states gradually increases as we increase the $g$ value. The whole observation is shown in Fig.6.

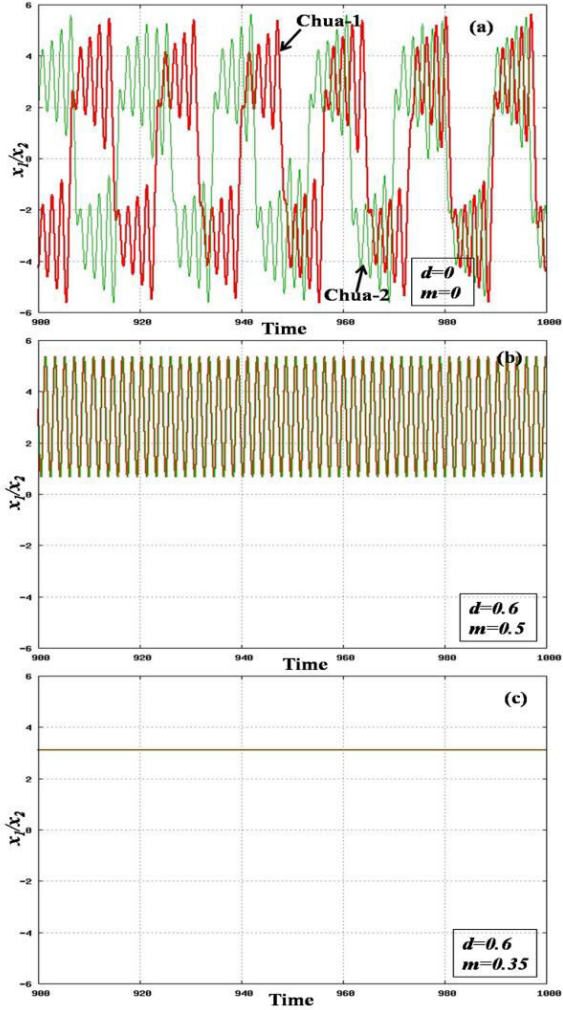

FIG.4. (Color online) Amplitude death (AD) in identical Chua's circuit: (a) the chaotic dynamics in two isolated Chua's circuits; (b) oscillation of period-1 dynamics of MFD coupled Chua circuit and (c) occurrence of AD in the coupled system. The other system parameters are: . $\alpha_1 = \alpha_2 = \alpha = 10$, $\beta_1 = \beta_2 = \beta = 18.43$, $\gamma = 1$, $d_1 = d_2 = d$ and $g_1 = g_2 = g = 0$, respectively. AD is achieved in the coupled oscillators for $d = .6$ and $m = 0.35$.

However, it is practically impossible to construct identical oscillators in experiments, because of the tolerance values of the circuit components. Therefore, a study with non-identical parameters is also carried out. For a typical set of parameter values where $\alpha_1 = 10$, $\alpha_2 = 10.5$, $\beta_1 = 18.63$, $\beta_2 = 17.5$, $\gamma = 1.05$, $m = 0.35$ both the system enter into the AD region with higher diffusive strength i.e. $d_1 = 0.8$ and $d_2 = 0.76$. In this condition also with a very low value of $g$ (i.e. $g = 0.015$) OD appears through symmetry breaking. Note that, if we apply different $g$ values for two circuits (which is possible by differing $r_{11}$ and $r_{21}$) the OD appears through asymmetry breaking [Not shown through figure].

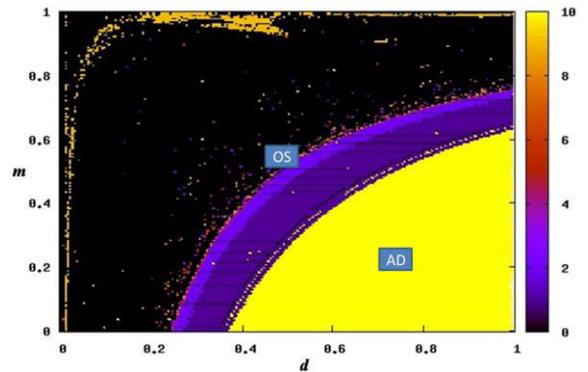

FIG.5. (Color online) Transition from the region of oscillatory state (OS) to amplitude death (AD) as observed through numerical simulation in parameter space $(d - m)$ for two identical Chua's circuits coupled under

MFD coupling. The other system parameters are: $\alpha_1 = \alpha_2 = \alpha = 10$, $\beta_1 = \beta_2 = \beta = 18.43$, $\gamma = 1$, $d_1 = d_2 = d$ and $g_1 = g_2 = g = 0$, respectively. Here, OS defines the oscillatory dynamics with different periods (0-9 in color bar) and AD (10 in color bar) the amplitude death state.

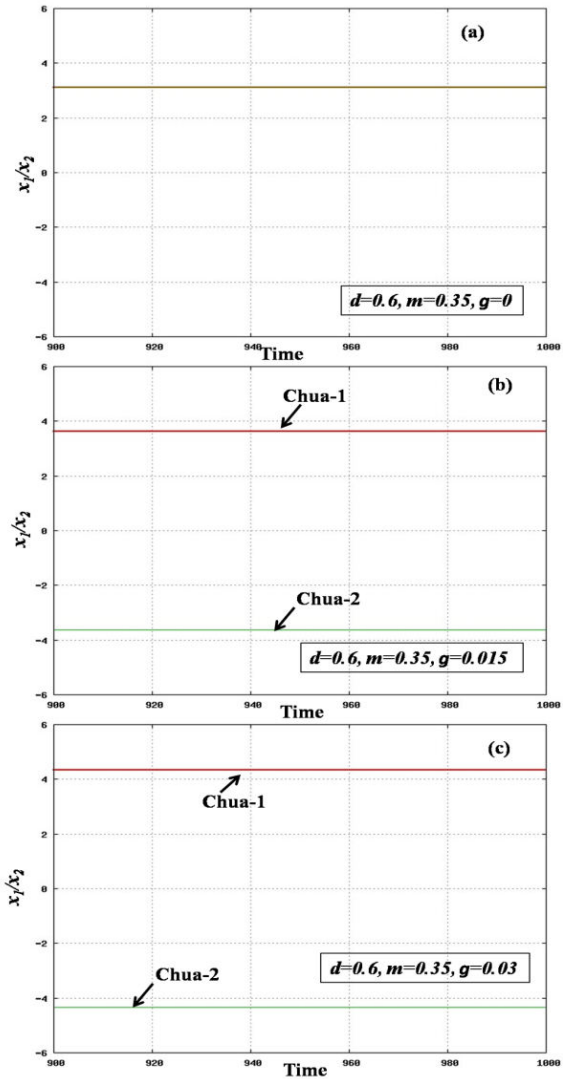

FIG.6. (Color online) Transition from the amplitude death (AD) to Oscillation death (OD) as observed through numerical time series plots of $x_1$ and $x_2$ for two identical Chua's circuits coupled through MFD coupling as well as direct coupling. The other system parameters are: $\alpha_1 = \alpha_2 = \alpha = 10$, $\beta_1 = \beta_2 = \beta = 18.43$, $\gamma = 1$, $d_1 = d_2 = d$ and $g_1 = g_2 = g = 0$, respectively. Here, (a) AD state for $g = 0$, $d = 0.6$ and $m = 0.35$; (b) OD is achieved for $g = 0.015$, and (c) separation between two steady states increase for $g = .0.03$.

## IV. EXPERIMENTAL RESULTS

Hardware circuits of the MFD as well as direct coupled Chua circuit as shown in Fig. 3 are designed on a bread board using IC TL082 (op-amp for Chua diode, inductor block and MFD coupling), capacitors and resistors etc. Here, we use a $\pm 12$ volt power supply. The two Chua's circuits are designed using the following parameters [20]: $R_{11} = R_{21} = 220$ ohm, $R_{12} = R_{22} = 220$ ohm, $R_{13} = R_{23} = 2.2$ kohm, $R_{14} = R_{24} = 22$ kohm, $R_{15} = R_{25} = 22$ kohm $R_{16} = R_{26} = 3.3$ kohm, $R_{17} = R_{27} = 100$ ohm, $R_{18} = R_{28} = R_{19} = R_{29} = 1$ kohm, $R_{110} = R_{210} = 2.2$ kohm, $C_{12} = C_{13} = C_{22} = C_{23} = 0.1 \mu F$, $C_{11} = C_{21} = 0.01 \mu F$ and 2 k POT for each $r_{11}$ and $r_{21}$. Here, the inductors in Chua's circuits are replaced by general impedance converters [23]. The effective inductances of general impedance converters are 22 mH each. The MFD coupling block has been constructed with $R_1 = R_2 = R_4 = R_5 = 10$ kohm and $R_3 (\propto m)$, $R_6 (\propto \frac{1}{d})$ with 10 k POT. The direct coupling has been done by using a high value resistor $R_7$.

To explore the dynamics of such modified circuit we have first set $r_{11} = 1.84$ kohm and $r_{21} = 1.72$ kohm. With these values both the system show chaotic dynamics (Fig.7(a)). Then we apply MFD coupling. When we set $R_3 = 1.59$ kohm (i.e. $m = 0.32$) and $R_6 = 2.35$ kohm (i.e. $d \approx 0.78$) the chaotic dynamics converges to oscillation of period-1. At $R_3 = 1.2$ kohm (i.e. $m = 0.24$) the oscillatory behavior of both the systems completely vanishes and they converge to the same steady state (single dc line appears on the oscilloscope). Thus, AD occurs. These two results are shown in Fig.7(b) and Fig.7(c), respectively. In all these cases the direct coupling connection is taken as open (i.e. $R_7 = \infty$).

Next we experimentally verify the influence of $R_7$ ($\propto \frac{1}{g}$) on the circuit. The AD state (single dc line) splits into two new steady states (two dc lines appear on the oscilloscope) even when we connect $R_7$ with a very high value (300 kohm). Thus OD appears. Now when we gradually decrease the $R_7$, the separation between the two steady states increases. Thus it supports our numerical predictions. The whole observation is shown in Fig. 8.

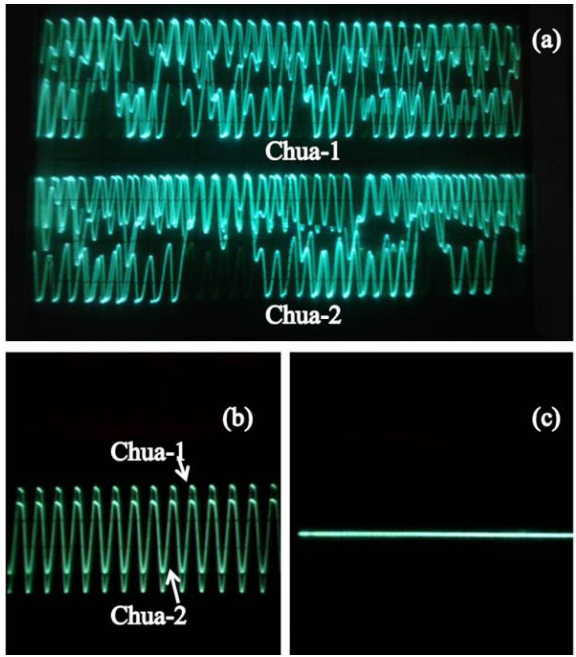

FIG.7. (Color online) Experimentally observed amplitude death (AD) in Chua's circuit. The time series traces of $v_{11}$ and $v_{21}$; (a) the chaotic dynamics in two isolated Chua's circuits for $r_{11} = 1.84$ kohm and $r_{21} = 1.72$ kohm; (b) oscillation of period-1 dynamics of MFD coupled Chua circuit for $r_{11} = 1.84$ kohm, $r_{21} = 1.72$ kohm, $R_3 = 1.59$ kohm and $R_6 = 2.35$ kohm and (c) occurrence of AD in the coupled system for $r_{11} = 1.84$ kohm, $r_{21} = 1.72$ kohm, $R_3 = 1.2$ kohm and $R_6 = 2.35$ kohm. The direct coupling connection is taken as open (i.e. $R_7 = \infty$).

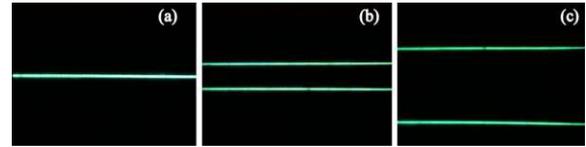

FIG.8. (Color online) Experimentally observed transition from the amplitude death (AD) to oscillation death (OD) for two Chua's circuits coupled through MFD coupling as well as direct coupling. The time series traces of $v_{11}$ and $v_{21}$; (a) the AD in coupled Chua's circuits for $r_{11} = 1.84$ kohm, $r_{21} = 1.72$ kohm, $R_3 = 1.2$ kohm, $R_6 = 2.35$ kohm and $R_7 = \infty$ ; (b) occurrence of OD for $R_7 = 300$ kohm; and (c) separation between two dc line increase for $R_7 = 60$ kohm.

## V. CONCLUSION

We have explored the phenomena of transition from AD to OD in the MFD as well as direct coupled Chua's circuits. Using detailed numerical simulations we have shown that an additional direct coupling can induce OD in a MFD coupled Chua's circuit and a transition between AD to OD occurs. Numerically we also show that the fact is true for both identical and non-identical systems. It has been shown that OD appears in the system even when the direct coupling is very weak. We explore that using different direct coupling strength one can easily creates different OD states. According to our knowledge, this transition in Chua's circuit has not been observed earlier. We have also experimentally observed this transition on a prototype hardware circuit and all the experimental results support our numerical predictions. This study includes a new dimension to the study of versatile nonlinear phenomena that one can find in most simple Chua's circuit. This study can be extended to other chaotic systems to improve our understanding. We also hopeful apart from electronic circuits, such AD-OD transition can be observed in engineering and biological systems and it may reveal the practical application of this transition.

## ACKNOWLEDGMENTS
The author acknowledges the infrastructural support from the Physics Department, Bidhan Chandra College, Asansol- 713304.